\documentclass[conference]{IEEEtran}

\usepackage[english]{babel}
\usepackage{t1enc}
\usepackage{cite}
\usepackage{url}
\usepackage[T1]{fontenc}
\usepackage[utf8]{inputenc}
\usepackage{moreverb}

\usepackage{setspace}
\usepackage{listings,multicol}
\usepackage[table,usenames,dvipsnames,svgnames]{xcolor} 
\usepackage{epstopdf}
\usepackage{graphicx}
\usepackage{enumitem}
\usepackage{booktabs}
\usepackage{textcomp}
\usepackage{diagbox}
\usepackage{footnote}

\usepackage[backgroundcolor=blue!10,bordercolor=Gray]{todonotes}
\usepackage{fancyhdr}

\usepackage{graphicx}
\graphicspath{ {resources/} }

\definecolor{listinggray}{gray}{0.9}
\definecolor{lbcolor}{rgb}{0.9,0.9,0.9}
\begin{document}

%\selectlanguage{magyar}

%\title{C++11 dialektus használatának támogatása örökölt környezetben}
%\title{Supporting the C++11 Dialect in Legacy C++ Development Environments}
%\title{Supporting the C++11 Dialect in Legacy C++ Compilation Environments}
\title{Transforming C++11 Code to C++03 to Support Legacy Compilation Environments}

% author names and affiliations
% use a multiple column layout for up to three different
% affiliations
%\author{\IEEEauthorblockN{Michael Shell}
%\IEEEauthorblockA{School of Electrical and\\Computer Engineering\\
%Georgia Institute of Technology\\
%Atlanta, Georgia 30332--0250\\
%Email: http://www.michaelshell.org/contact.html}
%\and
%\IEEEauthorblockN{Homer Simpson}
%\IEEEauthorblockA{Twentieth Century Fox\\
%Springfield, USA\\
%Email: homer@thesimpsons.com}
%\and
%\IEEEauthorblockN{James Kirk\\ and Montgomery Scott}
%\IEEEauthorblockA{Starfleet Academy\\
%San Francisco, California 96678--2391\\
%Telephone: (800) 555--1212\\
%Fax: (888) 555--1212}}

\author{\IEEEauthorblockN{G\'abor Antal,
D\'avid Havas,
Istv\'an Siket,
\'Arp\'ad Besz\'edes,
Rudolf Ferenc}
\IEEEauthorblockA{Department of Software Engineering\\University of Szeged, Szeged, Hungary}
\IEEEauthorblockA{\{antal,havasd,siket,beszedes,ferenc\}@inf.u-szeged.hu}
\and
\IEEEauthorblockN{J\'ozsef Mihalicza}\\
\IEEEauthorblockA{NNG LLC}
\IEEEauthorblockA{jmihalicza@gmail.com}
}

% make the title area
\maketitle

% As a general rule, do not put math, special symbols or citations
% in the abstract

\begin{comment}
Newer technologies -- programming languages, environments, libraries -- change very rapidly.
Unfortunately, not all projects are able to quickly adopt to the changes due to various internal and external constraints.
Such a situation is when the software vendor delivers the product to a customer who requires specific platform compatibility.
In this work, we deal with this issue in the context of the C++ programming language.
Our industrial partner is in a situation that only older C++ language editions can be used in production code, however it would like to give the possibility to its developers to use the newest constructs in their development environment.
To aid this problem, we created a source code transformation framework with which C++ source code written according to the C++11 standard can be automatically backported to C++ code conforming to C++03.
This way, the developers are writing their code in the development environment exploiting the latest language features, while production code is built using only the allowed language constructs.
This paper reports on the technical details of the transformation engine, and our experiences in applying it on two large industrial code bases and four open source systems.
\end{comment}

\begin{abstract}
Newer technologies -- programming languages, environments, libraries -- change very rapidly.
However, various internal and external constraints often prevent projects from quickly adopting to these changes.
Customers may require specific platform compatibility from a software vendor, for example.
In this work, we deal with such an issue in the context of the C++ programming language.
Our industrial partner is required to use SDKs that support only older C++ language editions.
They, however, would like to allow their developers to use the newest language constructs in their code.
To address this problem, we created a source code transformation framework to automatically backport source code written according to the C++11 standard to its functionally equivalent C++03 variant.
With our framework developers are free to exploit the latest language features, while production code is still built by using a restricted set of available language constructs.
This paper reports on the technical details of the transformation engine, and our experiences in applying it on two large industrial code bases and four open-source systems.
Our solution is freely available and open-source.
\end{abstract}

% no keywords


% For peer review papers, you can put extra information on the cover
% page as needed:
% \ifCLASSOPTIONpeerreview
% \begin{center} \bfseries EDICS Category: 3-BBND \end{center}
% \fi
%
% For peerreview papers, this IEEEtran command inserts a page break and
% creates the second title. It will be ignored for other modes.
\IEEEpeerreviewmaketitle

\section{Introduction}
\label{chap:intro}

%\todoi{1 oldal, Arpi}

Today, technologies used in software engineering practice, such as programming languages, environments and libraries, change on an unexperienced pace.
And, naturally, developers would like to exploit the advantages of such developments in order to increase their productivity, quality of code and reduce risks of error.
However, often there are certain constraints in the projects that prohibit using the newest technologies.
This includes, for instance, interoperability with legacy systems, compatibility with older hardware and software, and other limitations arising from the context of the project.
For instance, in a situation when the software vendor delivers software to a customer, it must conform to the customer's requirements regarding platform compatibility.

The work presented in this paper was motivated exactly by such a situation.
NNG LLC, our industrial partner, is a company that develops navigation software, and as such it delivers software products to its clients who integrate the navigation software component into the host
system of the final product.
These host systems often raise strict technical constraints against the delivered software to be integrated.
Compatibility may be required with old operating systems, libraries, and existing components.
Consequently, the development company needs to enforce strict regulations in-house regarding the usable platforms, language versions and development environments.
The net effect is that the developers are confronted with a situation in which they are limited by older technologies, while they would be eager to use more advanced ones.
Often, this leads to lower productivity and even lack of motivation because their professional skills development is limited as well.

In this work, we deal with the mentioned problems in the context of the C++ language, the primary technology used by the company.
For many years, the official language standard has not been updated until 2011, which progressively resulted in the birth of a large code base globally, which is now treated already as legacy code.
The C++11 standard~\cite{cpp:standard11} included so many new features (such as in-class initializations, lambda functions, automatic types, attributes, and many more) that made it almost a new language (even Bjarne Stroustrup, the creator of C++ thinks it ``feels like a new language\footnote{\url{http://www.stroustrup.com/C++11FAQ.html\#think}}'').
However, even after five years of the publication of the new standard, developers at NNG are still forced to use older versions of the language, which is a significant drawback from both the subject system and from the developers' point of view.

Hence, the goal of our R\&D collaboration project was to develop a solution to this problem in a way that would be both beneficial for the developers and the system itself.
We created a source code transformation framework with which C++ source code written according to the C++11 standard can be automatically ``backported'' to C++ code conforming to earlier language versions (C++03, in particular~\cite{cpp:standard03}).
The framework is capable of automatically transforming a large number of new language constructs to their equivalent versions in the older language.
This way, developers are free to exploit the latest language features, while production code is still built by using a restricted set of available language constructs.
Even though various technical limitations prevented us from making a complete transformation solution in terms of supported language elements,
our framework enables a very large subset of C++11, making it usable in practice.

The transformation framework includes a number of additional features besides transforming individual source code files, which make its integration into practical build processes easier.
These include, among others, source tree mirroring, incremental transformation, selective transformation, and traceability between the original and the transformed code.
The technology has been experimentally integrated into the development process of the company (which was not trivial due to some unique properties of the build process), enabling them
to benefit from using recent technology while retaining compatibility with their partners using legacy systems.

This paper reports on the technical details of the transformation engine, and our experiences in applying it not only on NNG's code base but on another industrial application and on four open source systems as well.
Although the transformations do not cover C++11 in 100\%, our results and experiences with industrial systems indicated that in its present state the framework is definitely useful in practice.
The transformation engine is available open-source:\newline
\texttt{\url{https://github.com/sed-szeged/cppbackport}}

The paper is organized as follows.
Section~\ref{chap:motiv} presents more details on the practical scenario that lead to the development of the solution.
Related work is briefly presented in Section~\ref{chap:related}.
Section~\ref{chap:framework} describes the framework and its usage scenarios in detail, while the transformations themselves are listed in Section~\ref{chap:trafo}.
Section~\ref{chap:eval} deals with the evaluation of the solution and our measurement results, together with
Section~\ref{chap:limit}, which lists the most important limitations of the approach,
before the conclusion in Section~\ref{chap:concl}.

\section{Motivation and Overview}
\label{chap:motiv}

iGO navigation software, the core product of NNG, is a {\em white label} product, meaning that clients can sell the final products under their own brand.
Clients have significant freedom in customizing the user interface and application behavior to their taste, which produces high variability not only on the market,
but on a technical level as well.
While customizations have big impact on certain features and workflows, many core functionalities remain practically the same in the majority of the products.
As a typical software product line \cite{Pohl:2005:SPL:1095605}, the iGO system has core assets that share a common code base, which has to compile in all supported environments.

In some segments, successful products have numerous new generations with newer and newer versions of the iGO core in them, but without significant changes in the hardware/OS layers.
iGO core assets are required to support compilation environments for these legacy platforms as long as business interest \cite{363157} and support periods sustain the need.
Two notable examples of such legacy target platforms are Windows CE and QNX 6.5. Windows CE can only be targeted with C++03 compilers, while for QNX 6.5 the compilation toolchain is based on GCC~4.4.2.

On one hand we see a clearly articulated C++03 compatibility requirement for several years.
On the other hand C++11 and the more recent versions of the C++ language are not only minor refinements, but contain significant benefits over the legacy language.
There are multiple aspects here. One group of them relates to product quality. Move semantics of C++11 allows faster code even without modifying the source code~\cite{Meyers:2014:EMC:2685398}.
Many features of the new language help to enhance code expressiveness. Self-explanatory code without boilerplates is less error-prone and in turn leads to better quality and faster production.

The other key factor is developer retention/attraction. Not having major changes to C++98, in a few years we can refer to C++03 as a 20-years technology.
Continuous learning is a vital part of the successful developer mindset \cite{Martin:2011:CCC:1999258}.
Reliable extension and maintenance of a multiplatform C++ software product line requires skilled engineers, for whom modern C++ is the norm.
Being forced to a 20-years technology with millions of lines of code in a non-trivial domain easily becomes a business issue because of this human factor.

The opposing business needs for the legacy and new C++ variants made NNG think in building a bridge between the two.
The requirement is simple: be able to use as many of the modern C++ features in the common code base as possible without compromising compatibility with the still important legacy platforms.
% \todoi{kb. itt lehet vágni}

Our first cooperation in this topic was a classic research project to come up with possible approaches and their detailed assessment for decision making.
Table~\ref{tbl:assessment} contains the identified scenarios and their fitness from different angles.
The three possibilities were: {\em Columbus}, a C++ analysis framework developed at the University of Szeged~\cite{FBT02}, the open-source {\em clang} front end for the LLVM infrastructure~\cite{comp:clang}, and the {\em C backend} developed also for LLVM~\cite{s2s:cbe}.
Each criterion was assessed on a scale of 1--5, as can be seen in the table.
%Clang AST and Columbus are explicit code transformations using the clang or the proprietary Columbus compilation frontend respectively.
%C backend is the revival of the abandoned llvm $\rightarrow$ C backend.
Finally, NNG decided to choose the \emph{clang} code transformation approach, mostly because it is open-source while Columbus is not, and the C backend turned out to be incomplete and unreliable.

\begin{table}[htb]
\centering
\caption{Possible transformation scenarios and their fitness (1~-~bad, 5~-~good) from different angles.}
\label{tbl:assessment}
\begin{tabular}{|p{3.9cm}|p{0.75cm}|p{1.35cm}|p{1.0cm}|}
\hline
Criterion & LLVM & Columbus & LLVM  \\
& clang & & C back \\
\hline
Cost of development & 2 & 2 & 1 \\
\hline
Cost of integration into NNG processes & 4 & 4 & 3 \\
\hline
Learning curve & 5 & 5 & 3 \\
\hline
Degradation of work efficiency & 3 & 3 & 1 \\
\hline
Diagnostics & 4 & 4 & 1 \\
\hline
Performance: compilation & 2 & 1 & 1 \\
\hline
Performance: speed & 4 & 4 & 1 \\
\hline
Performance: memory & 4 & 4 & 3 \\
\hline
Performance: executable size & 4 & 4 & 3 \\
\hline
New language elements & 1 & 1 & 4 \\
\hline
Robustness & 3 & 3 & 5 \\
\hline
Future proof & 1 & 1 & 3 \\
\hline
Automation & 5 & 5 & 5 \\
\hline
Impact on iGO code & 5 & 5 & 5 \\
\hline
Support & 3 & 4 & 1 \\
\hline
Legacy compatibility & 5 & 5 & 5 \\
\hline
\end{tabular}
\end{table}

\subsection{Overview of the solution}
\label{chap:motiv:over}

\begin{figure}[htb]
\centering
\includegraphics[width=\linewidth]{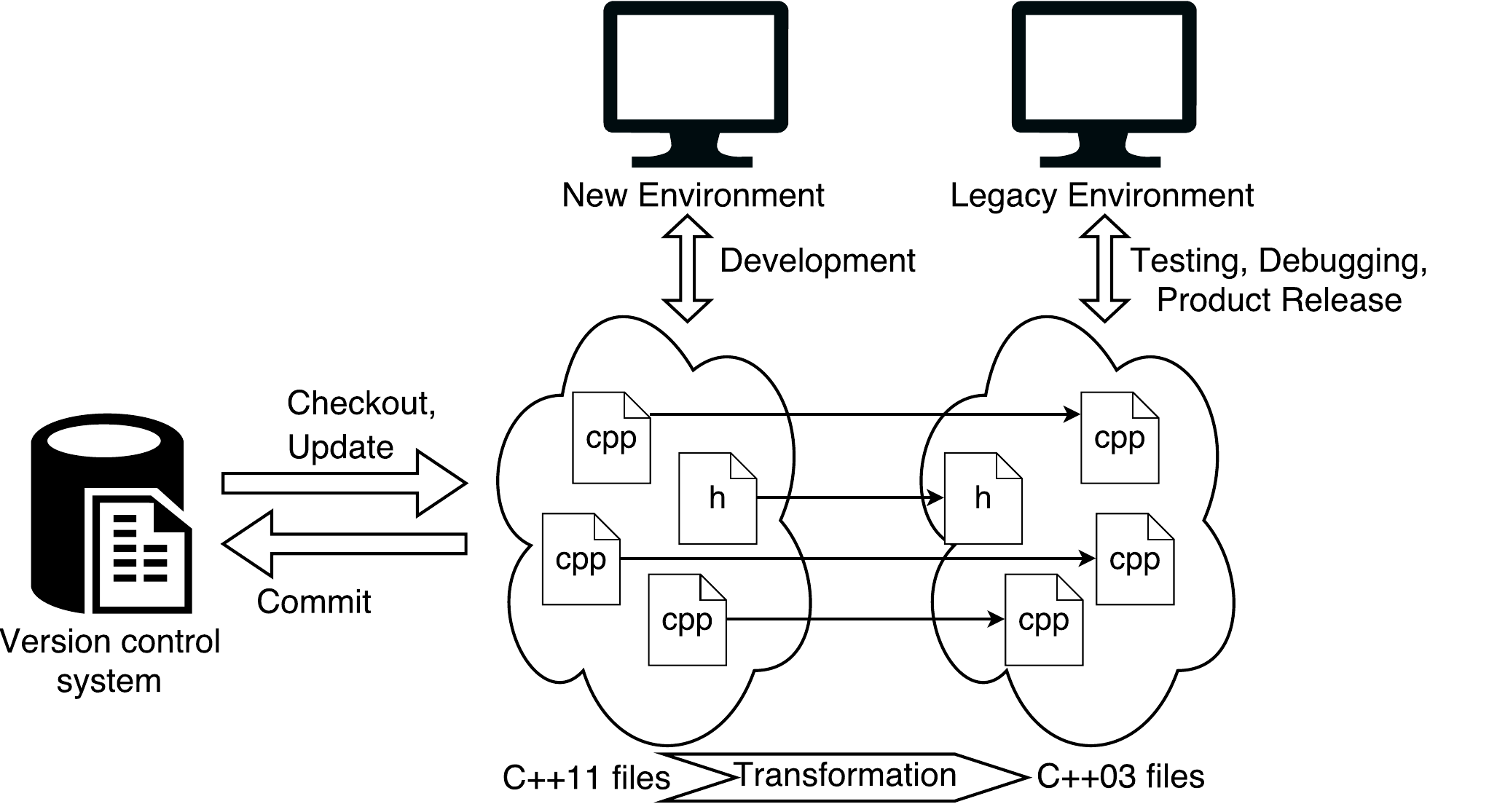}
\caption{General use case of the framework}
\label{fig:intro}
\end{figure}

A high-level overview of the transformation process is depicted in Figure~\ref{fig:intro}.
Developers use a modern C++ IDE (e.g. Microsoft Visual Studio\footnote{https://www.visualstudio.com} 2015) in their daily job.
Our tool generates the backported equivalent of the source tree, so when a legacy build or debugging is needed, legacy tools/IDEs (e.g. Microsoft Visual Studio 2005 or GCC 4.4.2) can be used naturally.

Apart from the transformation itself, our framework provides support for various every day software engineering activities such as testing and debugging.
Since runtime issues (either from testing or operation phases) arise at the legacy production environment, while the developers should use their native development environment, the necessary traceability needs to be established on source code level.

For instance, bug reports of native systems may contain location references to the compiled executable. In case of a crash, for example, call stacks of different threads are dumped.
This information together with a corresponding map file that matches the raw addresses to the source code are invaluable for finding the root cause of the bug.
On legacy targets call stacks refer to the backported source code. For more seamless integration into the development processes, we have created a convenience tool that enables developers to lookup
the source code location in the modern C++ source code even for addresses referring to the backported executable.

\section{Related Work}
\label{chap:related}

This work deals with static code analysis for the purpose of source-to-source code transformation.
The topic has a large literature, and there are many experimental and production tools developed for various languages, both free and commercial.
Also, the application areas are diverse: language translation, (back)porting, modernization, refactoring, etc.
In this section, we overview the common solutions for source transformation with special focus on the C++ language, and not particularly on the application of transformation.

%Legacy systems written in languages like Cobol, Fortran, or even C and C++ are often the subject of source transformation to more modern languages like Java (see, for instance,~\cite{Bodin94sage++:an,harrycobol,egyeb}).

Compiler infrastructures are often used for language translation, for instance the EDG front end~\cite{comp:edg}, GNU GCC~\cite{comp:gcc}, the ROSE compiler infrastructure~\cite{comp:rose} and LLVM clang~\cite{comp:clang}, which is the chosen platform for our tool as well.
%A C\# programozási nyelvben hasonló lehetőségeket biztosít a \textit{Roslyn}~\cite{comp:roslyn} fordító.

\begin{comment}
\textit{Szőke és társai}~\cite{szoke2015faultbuster} egy olyan automatikus refaktoráló eszközt mutattak be, amely túlmutat a fejlesztőkörnyezet által kínált megoldásokon. Az alkalmazás saját felülettel is rendelkezik, de integrálható bizonyos fejlesztőkörnyezetekbe (pl.: Eclipse, NetBeans, IntelliJ). 
A FaultBuster nevű eszközük segítségével többféle statikus kódolási hibát, másnéven szabálysértést lehet refaktorálni, mint például üres catch blokkok, túl hosszú sorok, hosszú osztályok tördelése erősen összefügő komponensek mentén. Ez az eszköz a kézzel irányított automatikus refaktoring mellett képes kötegelve is végrehajtani transzformációkat, ám ekkor csak egy konkrét kódolási hiba javítható egyidőben.

Korlátoltabb lehetőségeket biztosít az \textit{Eclipse CDT}~\cite{comp:ecdt}, ami az Eclipse fejlesztőkörnyezethez ad C++ támogatást.
Segítségével változó átnevezéseket, függvény kiszervezéseket, szintaxis ellenőrzést, egyéb hibák ellenőrzését hajthatjuk végre.
Továbbá támogatja a kódformázást és a függvény kiszervezést, valamint fejlesztőkörnyezetbe integrálva lehetőséget ad szintaxis kiemelésre és automatikus kiegészítésre.
Hasonló megoldásokkal találkozhatunk más nyelvek esetén is.
\end{comment}

There are solutions that not only offer a library for source transformation but a complete framework for this task.
These frameworks often provide an own language to define the transformation and are easier to use being specific, though often bring higher overhead, more difficult learnability and less flexibility.
For example, Lee et al.~\cite{Lee2004} created such an environment, which is highly flexible and can be extended with new languages as well.
A similar system was offered by Bagge et al.~\cite{Bagge01codeboost:a} that provides support for source code instrumentation and optimization transformations, but this system supports only C++.
There are additional experimental and commercial systems which could be possibly suitable for similar tasks, such as SrcML~\cite{collard2010lightweight}, TXL~\cite{comp:txl}, ASF+SDF~\cite{comp:asfsdf}, Stratego~\cite{comp:stratego}, DMS toolkit~\cite{comp:dms}, and several others.

%\subsection{Transzformációs alkalmazások}

We found that only LLVM provides a proper interface to its internal representation that is suitable for our purposes, so we are using this environment.
A few additional applications based on the LLVM clang~\cite{comp:clang} front end are listed below.
Clang Tools~\cite{s2s:clangtools} is a toolset that includes a code transformation module as well.
An interesting tool is modernizer, which transforms C++03 code to C++11, exactly the opposite of what we developed.
This tool is appropriate for other tasks as well such as formatting and code style checking.
Another application of this library is Include What You Use~\cite{s2s:iwyu}, with which the optimization of include files can be performed.

Transformation on C++ code for a different purpose was done by, for instance, Aigner et al.~\cite{Aigner1996}, which can be used to eliminate virtual function calls in C++ in order to improve the performance of the programs.
Marangoni et al.~\cite{Marangoni:2016-02-14T00:00:00:2470-1173:1} implemented a tool with which general C++ code can be automatically transformed to CUDA source code, which enables parallel execution of general C++ on video cards.
Additional parellelization transformation tools have been implemented by Krzikalla et al.~\cite{Krzikalla2012} and Magni et al.~\cite{Magni:2013:LCE:2503210.2503268}.

An interesting tool based on LLVM is C Backend~\cite{s2s:cbe}, which is able to transform C++ code to C code.
This could have potentially also been a solution to our problem (as most compilers still support C), however this system is still in a very experimental phase.
The generated code is much slower than the original, furthermore it cannot handle a number of code constructs at all.

\section{Source Code Transformation Framework}
\label{chap:framework}

%\todoi{2 oldal, Rudi}

%A forráskód átalakítás folyamatát a transzformációs keretrendszer vezérli.
The alteration of the source code is controlled by the transformation framework.
%%FR A keretrendszer kialakításakor szerettünk volna egy ,,biztos alapot'' amire építhetünk, így a keretrendszer tervezésekor több, már meglévő rendszert is megvizsgáltunk és ezek alapján választottunk egy olyan architektúrát, ami minden, általunk támasztott követelménynek megfelel (azaz támogatást nyújt kódtranszformációhoz, valamint kódgeneráláshoz).
%%FR A vizsgált rendszerek közé tartozik a Microsoft Visual Studio fordítóarchitektúrája, amely semmilyen szinten nem biztosít forráskód transzformációhoz szükséges interfészt. 
%%Emellett megvizsgáltuk a Columbus technológia\footnote{https://www.sourcemeter.com} nyújtotta lehetőségeket is, amely támogatta ugyan a transzformációhoz szükséges műveleteket, viszont a kódgenerálásra nem nyújt megoldást. 
%%FR Megvizsgáltuk továbbá clang-llvm architektúrát is, amely minden szükséges igényünket teljesítette, azaz segítséget nyújtott a transzformációkhoz is, valamint a kódgeneráláshoz is, így erre az architektúrára esett a választásunk, a keretrendszer implementációjakor erre a rendszerre támaszkodtunk. 
%A keretrendszer két főbb részből áll: az első az inkrementális transzformálást biztosító rendszer, a második pedig a transzformációk végrehajtásáért felel.
It consist of two main parts: the first one is the engine providing incrementality, while the second one is responsible for performing the actual transformations.
%Az inkrementalitást biztosító rendszer a fájl szintű változások követése által meghatározza, hogy a projekt mely fájljait kell transzformálni, majd ezen lista alapján a transzformációs rendszer elvégzi a tényleges átalakítást.
The incrementality engine monitors the code changes at file level and determines which files of the project need to be transformed (discussed in more detail in Section~\ref{sec:increm}).
Based on this list, the transformation engine performs the needed changes, which is the topic of Section~\ref{chap:framework:trans}.

%A keretrendszer tervezésekor kiemelt szempont volt, hogy az alkalmazás könnyen integrálható legyen a fejlesztési folyamatokba; akár fordítás előtti lépésként (pre-build step), akár pedig CI\footnote{https://en.wikipedia.org/wiki/Continuous\_integration} (folyamatos integrációt biztosító) rendszerekben, így az implementációkor erre különös figyelmet fordítottunk.
During the design of the framework, it was an important requirement that the tool should be easy to integrate into the build processes; either as a pre-build step in traditional build systems or into continuous integration (CI) environments.
%\footnote{https://en.wikipedia.org/wiki/Continuous\_integration} systems.

\subsection{Source code transformation}
\label{chap:framework:trans}

%A transzformációs keretrendszer használatához tudnunk kell, hogy egyes fordítási egységeket hogy fordítják az eredeti környezetben (amikor a transzformációs keretrendszertől függetlenül szeretnénk binárisokat előállítani a forráskódból). 
For using the transformation framework, we have to know how the compilation units are compiled in their original build environment.
%Ehhez a clang által is használt \texttt{compile\_commands.json} fájlt \cite{clang:compilecommands} használjuk. 
We use the \texttt{compile\_commands.json} file~\cite{clang:compilecommands} for this purpose.
%Ebben a szöveges fájlban, a JSON szintaktikai szabályai által meghatározott\footnote{Bővebben: http://www.w3schools.com/json/json\_syntax.asp} módon tároljuk a fordítási egységekhez szükséges adatokat, amik a következők:
This text file contains the necessary information, which is the following:

\begin{itemize}[noitemsep]
	%\item \textbf{directory} (könyvtár): A fordítás során használt munkakönyvtár. A további mezőket (command, file) az ebben megadott útvonalhoz képest relatívan adhatjuk meg. 
	\item \textbf{directory}: the working directory used during the build process. The following fields (command, file) are relative to this path.
	%\item \textbf{command} (parancs): A fordítási egység lefordításához használt utasítás.
	\item \textbf{command}: the command line used to compile the compilation unit.
	%\item \textbf{file} (fájl):  A fordítási egységhez tartozó fájl relatív, vagy abszolút útvonala.
	\item \textbf{file}: path of the compilation unit file.
\end{itemize}

%Ezeket az adatokat minden önálló fordítási egységhez meg kell adnunk.
This data has to be provided for each compilation unit.
%\Aref{lst:ccommands}. kódrészletben láthatunk egy példát a \texttt{compile\_commands.json} felépítésére.
In Figure~\ref{lst:ccommands} we show an example \texttt{compile\_commands.json} file content.
%Amennyiben egy projekt még nem rendelkezik ezzel a fájllal, akkor ezt a program felhasználójának kell elkészítenie.
If the project does not contain this file yet, then the user has to create one.
%A \texttt{compile\_commands.json} fájl elkészítése történhet automatikusan (külső eszköz, például CMake által), vagy pedig teljesen manuális módon.
The \texttt{compile\_commands.json} file can be created automatically (with an external tool, like CMake) or manually.
%Erre külön eszközt nem készítettünk, mivel az ipari partnerünktől sem érkezett igény erre, valamint egy garantáltan minden esetben jól működő eszköz elkészítése nem triviális feladat.
We did not prepare such a tool on our own, because the industrial partner did not require it.
\vspace{-3mm}

\begin{figure}[htb]
\begin{lstlisting}
[{
     "directory": "c:/work/projectDir", 
     "command": "cl.exe -c Source1.cpp -o2",
     "file": "c:/work/projectDir/Source1.cpp"
}]
\end{lstlisting}
\caption{Content of a \texttt{compile\_commands.json} file}\label{lst:ccommands}
\end{figure}

%A transzformációk megkezdése előtt a keretrendszer a teljes projekt hierarchiát átmásolja egy munkakönyvtárba, melyet a felhasználónak kell megadnia.
Before the transformation starts, the framework copies the full project hierarchy into a work directory, which has to be provided by the user.
%Ennek oka az, hogy a transzformált kód ebbe a mappába kerül, így a transzformációk befejezése után, a munkakönyvtárban lévő projektet lehet lefordítani a C++03 szabvány szerinti fordítóval.
The transformed code will be saved into this directory as well, so this code will be compilable with a C++03 compiler.
%%FR Viszont a forrásfájlokon kívül általában sok, egyéb fájl is megtalálható egy projekt mappájában, ilyenek lehetnek például a felhasználói felületen lévő képek, amelyek szükségesek a program futtatásához.

%A következőkben a forráskód transzformációs folyamat kerül bemutatásra.
In the following, we will describe the transformation process.
%A folyamat megtervezése során fontos szempont volt, hogy vannak olyan C++11 újdonságok, amelyeket nem lehet egy menetben átalakítani (például az egymásba ágyazott lambda kifejezés), illetve lehetnek olyanok is, amelyek egymásra épülnek, azaz több körben kell a transzformációt elvégezni.
During designing the process it was important to take into consideration that there are also such new C++11 features which cannot be transformed in one step (e.g. lambda expressions nested into other lambdas), and some transformations depend on each other and have to be performed in more iterations in a predefined sequence.

\begin{figure}[htb]
\centering
\includegraphics[width=\linewidth]{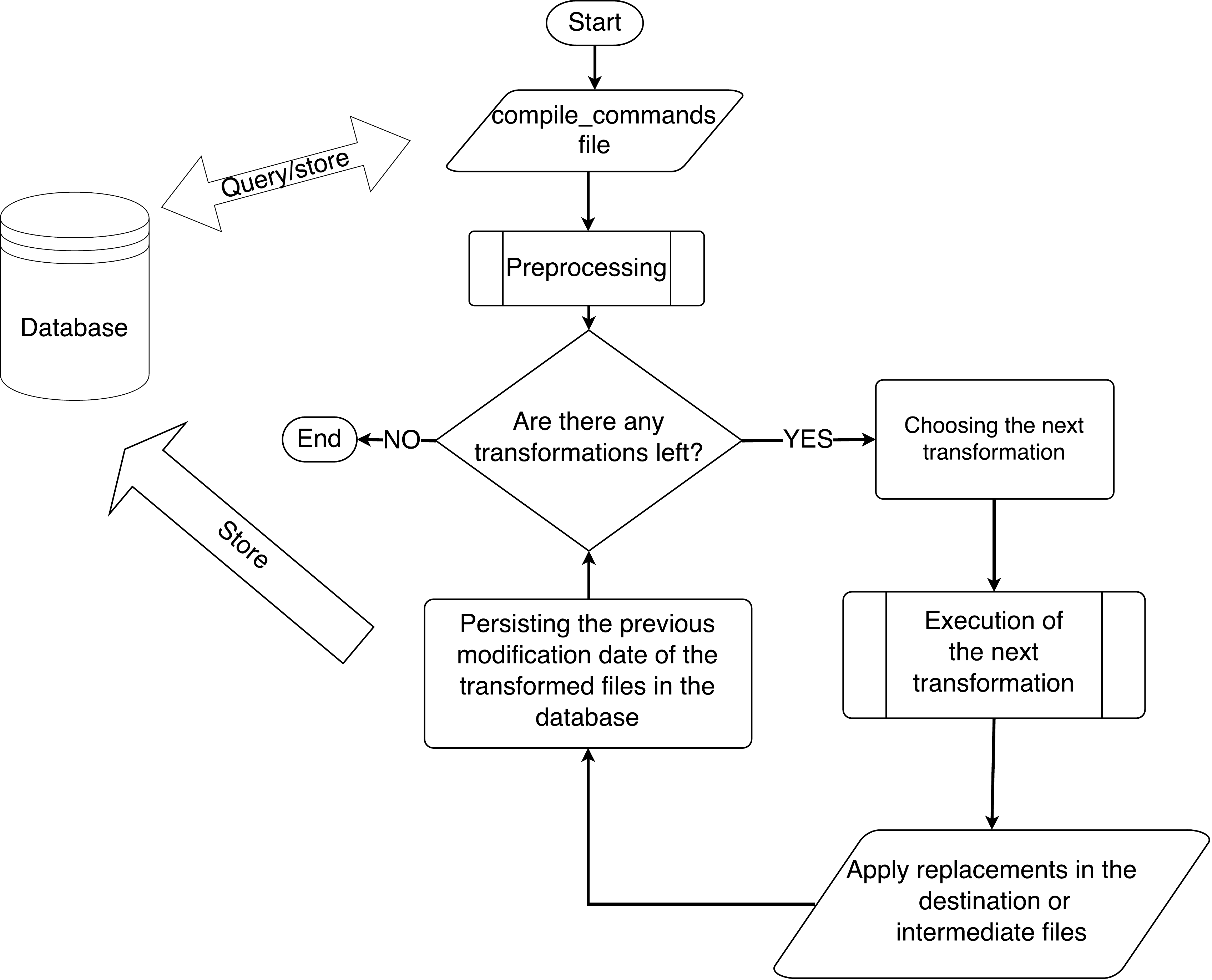}
\caption{Flow chart of the transformation framework}
\label{fig:trans_framework}
\end{figure}

%A transzformáció folyamatát mutatja be \aref{fig:trans_framework}. ábra, amelyben láthatóak az egyes fázisok. 
The transformation process and its phases are shown in Figure~\ref{fig:trans_framework}.
%Ezek a következők:
These are the following:

\begin{itemize}
	%\item A transzformációs eszköz a projekt fordítási adatait tartalmazó compile\_commands.json fájlt várja bemenetként.
	\item The transformation tool expects the \texttt{compile\_commands.json} file containing the project's compilation information as input.
	%\item Az előfeldolgozás során a transzformációs keretrendszer megvizsgálja a fordítási egységekhez tartozó függőségeket, és az adatbázis alapján kiszűri azokat a fájlokat, amiket transzformálnia kell.
	\item We maintain a database, which supports the incremental operation by storing the latest modification times and the dependencies between the source elements. During preprocessing, the transformation framework analyzes the dependencies between compilation units and selects those files which have to be transformed based on the database (see Section~\ref{sec:increm}).
	%\item Végighaladunk a transzformációk listáján. (A transzformációkról bővebben \aref{chap:trafo}. fejezetben írunk.)
	\item It then iterates over the list of the transformations. (We will describe these in Section~\ref{chap:trafo}.)
	%\item Miután egy transzformációt elvégeztünk az összes érintett fájlon, elmentjük a módosításokat, és az inkrementális rendszer frissíti az adatbázist, a transzformáció során érintett fájlok legutóbbi módosítási idejével.
	\item After a transformation is done on all affected files, the framework saves the changes, and the incrementality engine updates the database with the file modification dates.
\end{itemize}

\subsection{Incrementality}
\label{sec:increm}

%Rengeteg plusz munkát jelentene, ha minden futtatásakor minden egyes fájlt újra kellene transzformálnia az eszköznek.
It would take lots of resources to transform every file during each build of the project.
%Nyilvánvalóan felesleges lenne, hiszen általában - a fejlesztés során - a forráskódbázis csak egy töredékében végzünk módosításokat.
This would be superfluous in most cases, because usually only a small fraction of the code gets changed during a development iteration.
%Ennek kiküszöbölésére a transzformálni kívánt projekt forrásfájljairól nyilván kell tartanunk, hogy a kód melyik verzióját transzformáltuk már, hiszen a tényleges transzformációt csak abban az esetben érdemes végrehajtani, ha az adott fájl módosul.
To eliminate this overhead, the framework records for each compilation unit file which version of it was already transformed, and it performs the transformation only if the file was modified in the intervening time.
%Módosulás alatt értjük azt, amikor egy adott fájl legutolsó módosítási ideje megváltozik. 
A file is considered to be modified if its last modification time changed.
%Noha ez nem teljesen biztos megoldás, a gyakorlatban mégis elegendőnek bizonyul, a legritkább esetben fordul elő olyan, hogy a fájl legutolsó módosítási ideje változik, de a fájl tartalma ténylegesen nem.
This is not the perfect solution, as the time attribute of a file can change even if its content does not, but this happens quite rarely
and the side effect is not harmful.

%Mivel a fájlok adatait az alkalmazás két futása között is nyilván szeretnénk tartani, ezért szükséges ezen adatokat valamilyen perzisztens tárolóban eltárolnunk.
Because we need to preserve the information between consecutive runs of the framework, we store the data in a persistent storage.
%Ehhez az SQLite\footnote{https://www.sqlite.org/} nevű SQL-motort választottuk, mivel nem kell hozzá adatbázis szerver, és mindenféle konfiguráció nélkül, gyorsan beüzemelhető.
We chose the SQLite\footnote{https://www.sqlite.org/} SQL-engine, because it does not need a database server and can be used easily without any configuration.
%Ugyanakkor a transzformációs keretrendszer gyorsan átalakítható más adatbázis motorokkal (MySQL, PostreSQL) történő kommunikációra is, mivel a használt adatbázis-műveleteknek külön interfész (API) van definiálva.
However, the framework can be quickly adapted to other SQL engines (e.g. PostgreSQL, MySQL), if needed.

%\subsection{Relational Database Schema}
%\label{sec:rel_db}

%\todoi{az alabbi tabla leirasok roviden a kommetelt reszbol}

%Az adatbázisban tároljuk a fordítási egységeket (melyek a compile\_commands.json fájlban vannak definiálva), valamint az egyes egységekhez tartozó fájlokat. 
The database stores information about the compilation units (which are defined in the \texttt{compile\_commands.json} file) and associated files for each unit.
%A fordítási egységekhez tartozó fájlokról eltároljuk a fájlok legutóbbi módosítási idejét, a fájlokhoz tartozó függőségeket, valamint a függőségek felvételekor látott módosítási időt.
For each compilation unit, it stores the last modification date, the file dependencies (such as due to inclusion or given in command-line arguments), and the time stamp of the dependency addition.
%Amennyiben egy fordítási egység tartalmaz (include-ol) egy headert, amiben szintén szerepelnek include-ok, akkor ezek a függőségek közvetlenül az adott fordítási egységhez fognak tartozni, nem pedig egy függőséghez. 
If a translation unit includes a header file, which also contains include-s, these dependencies will be added directly to the compilation unit, rather than to a dependency.
%A függőségfájloknak íly módon nem lehetnek függőségeik.
(Dependent files cannot have dependencies this way.)
%Ezeket figyelembe véve a következő adatbázis sémát dolgoztuk ki: 
Taking this into account, we developed the following simple database schema:

\begin{itemize}[noitemsep,leftmargin=*]
\footnotesize
\item[] \texttt{COMPILATION\_UNIT(\underline{id}, timestamp, cmd\_args)}
\item[] \texttt{FILES(\underline{id}, path)}
\item[] \texttt{RELATIONS(\underline{file\_id}, \underline{dep\_id}, dependency\_timestamp)}
\end{itemize}

\subsection{Operation of the Transformation Framework}

%A keretrendszer összegyűjti a \texttt{compile\_commands.json} fájlból a fordítási egységeket, majd ezeken végigmenve összegyűjti a fájlokhoz tartozó függőségeket (közvetlen és közvetett függőségeket is).
The framework collects the compilation units from the \texttt{compile\_commands.json} file and by iterating over this list it collects also their dependencies (direct and indirect ones as well).
%Ezek után ezt összeveti az adatbázisból lekérhető listával.
Next, it compares this information with the database contents.
%Amennyiben egy fordítási egység az előző transzformáció óta 
If a compilation unit

\begin{itemize}[noitemsep]
%\item módosult,
\item changed,
%\item a parancssori kapcsolói módosultak,
\item its command line arguments changed,
%\item függőségei módosultak (első módosításig vizsgáljuk),
\item its dependencies changed,
%\item új függőség jött létre,
\item new dependency appeared, or
%\item meglévő függőség törlődött,
\item existing dependency disappeared,
\end{itemize}

%akkor a fordítási egység a függőségeivel együtt bekerül a transzformálandó fájlok közé.
\noindent then the compilation unit gets inserted into the list of files to be transformed together with its dependencies.

%Az így elkészült lista tartalmazni fogja az összes fájlt, amely módosulhatott az előző transzformáció óta, így ezekre végre fogja hajtani a transzformációkat.
This list will contain all files which might got modified since the last transformation, and the framework will perform the transformation of these files.
%Amennyiben az összes transzformációt sikerül végrehajtania, a keretrendszer frissíteni fogja az adatbázist.
%Ez alatt értjük a benne lévő fájlok utolsó módosítási idejének frissítését, valamint az esetlegesen új függőségek felvételét, illetve a megszűnő függőségek törlését.
If all transformations finish successfully, the framework updates the database by saving the new modification dates, adding possible new dependencies or deleting the disappearing ones.
%Továbbá, az új fordítási egységek is bekerülnek az adatbázisba, a függőségeikkel együtt.
Furthermore, if new compilation units were added, these will also be added to the database together with their dependencies.

\subsection{First analysis}

%Az első elemzés megkezdése előtt az alkalmazás létrehozza az adatbázisfájlt.
Before starting the first analysis, the framework creates the database.
%Amennyiben már létezik az adatbázisfájl, a program indulásakor nem kerül felülírásra.
If it already exists, it will not be overwritten.
%Ezek után szükség esetén létrehozzuk a táblákat (bővebben a \nameref{sec:rel_db} részben).
Next, the data tables will be created (if needed).

%Az első alkalommal az alkalmazás minden fájlt transzformálni fog, ami nagyobb projekt esetében időigényes lehet.
During the first run, the framework will transform all files, which can be time consuming in case of a larger project.
%A későbbiekben azonban - az inkrementalitás miatt - a rendszer azon fájljait fogja csak traszformálni, amik vagy amiknek a függősége változott.
Later however, because of the incrementality, only the changed files will be transformed.

\subsection{Tracing the transformed code back to the original one}

%A traceability tool egy kiegészítő eszköz a transzformációs keretrendszerhez, melynek célja, hogy a transzformált és az eredeti projekt között egyfajta leképezést teremtsen, azaz képesek legyünk a sorok visszavezetésére az eredeti és az átalakított fájlok között.
The traceability tool is a complementary tool for the transformation framework, which aims to create a mapping between the transformed and the original project, that is able to trace the lines between the original and the converted files.
%Ez hasznos olyan esetekben, amikor a transzformált kódban előjön egy hiba, amit természetesen az eredeti kódbázisban kell kijavítani.
This is useful in cases when the transformed code contains an error, which, of course, has to be fixed in the original code.
%Ha megkap egy átalakított fájlt és a kérdéses sor számát, akkor visszaadja azt, hogy az a sor hányadik sornak felel meg az eredeti fájlban. 
If it receives a transformed file and the line number in question, it returns the corresponding line in the original file.
%Az eszköz úgy lett elkészítve, hogy C kódból is meg lehessen hívni a következő API-t használva:
%The tool has been developed such that it can be called also from C code using the following API:
%\begin{lstlisting}[label={lst:traceability},frame=tlrb]{Name}
%int traceBack(const char* filename, int targetLine);
%\end{lstlisting}

%\textbf{Megszorítások}

%Az eszköz használata korlátolt abból a szempontból, hogy a pontos visszavezetést csak abban az esetben tudja végrehajtani, ha a kiválasztott sor transzformációkon kívül esik.
Using the tool is limited in the sense that the back tracing can only be performed if it does not fall into a transformed region of code.
%Ha egy transzformáción belüli sor lett kiválasztva, akkor az adott sort tartalmazó legkülső transzformáció kezdősorának a visszavezetését tudjuk megadni.
If a line inside a transformed code part has been selected, it returns the back trace of the starting line of the outermost transformation.
%Ennek a korlátozásnak az az oka, hogy bizonyos átalakításoknál a függvények törzsét egy a clang által biztosított eljárással kell kiíratnunk. 
The reason for this restriction is that in case of some transformations the body of the transformed functions has to be written out with a procedure provided by clang.
%A probléma az, hogy bár ez a kód funkcionálisan meg fog egyezni az eredetivel, de formázásban eltér. 
The problem is that while this code will be functionally the same as the original, it will differ in formatting.
%Legegyszerűbb példa talán, hogy a jobb átláthatóság miatt használt kommentek, üres sorok nem kerülnek kiírásra. 
Perhaps the simplest example is that comments and blank lines are not printed out.
%%FR Így a jelenleg használt megoldással a transzformált részen nem lehet nyomon követni, hogy az egyes sorok az eredeti fájl melyik sorának felelnek meg.
%\todoi{ez a bekezdes szamomra egy kicsit erthetetlen volt}

\section{Transformation Catalog}
\label{chap:trafo}

\newcommand{\codexformarrow}{\begin{minipage}{0.07\linewidth}\centering$\,\Rightarrow$\end{minipage}}

%\todoi{3 oldal, Rudi+Arpi}

%\todoi{Nem vilagos, hogy ez a teljes lista vagy csak kivonat. Ha az utobbi, akkor miert pont ezek vannak itt?}

In this section, we present the transformation details of the actual language elements supported by the framework.
There are some other transformations available as well, which are in experimental phase and are mentioned in Section~\ref{chap:limit}.

NNG's selection of which language elements to transform was based on their subjective usefullness/benefit judgement and the required efforts and complexity.

\begin{comment}
A támogatott nyelvi újdonságokat külön-külön tárgyaljuk, részletesen bemutatva a megvalósítást példákon keresztül.
Továbbá, az egyes nyelvi elemek használatát is ismertetjük, valamint a megvalósítás során bizonyos nyelvi elemek esetén bevezetett korlátozások és hiányosságok is felsorolásra kerülnek.
Ezekre a korlátozásokra azért volt szükség, mert a megvalósításuk vagy nagyon sok időt és erőforrást igényelt volna, vagy nem is lehetne teljesen általánosan megvalósítani a használt eszközök segítségével.
Végezetül az egyes nyelvi elemeket megvalósító transzformációk tesztelésére kidolgozott módszer kerül bemutatásra.
\end{comment}

\subsection{In-class data member initialization}
%\todoi{Arpi}

The possibility to initialize class (union, struct) data members directly within their declaration in the class body has been introduced in C++11.
This has the benefit that a data member which has a default value need not be initialized in each constructor but only once directly after its declaration.
Earlier, this was possible only for data members with the \texttt{const static} modifier.
The syntax for this construct is to use assignment operator or the brace initializer of the form \texttt{\{ {\em value} \}}.
The construct has a restriction that only one member of unions can be initialized this way.

\begin{figure}[htb]
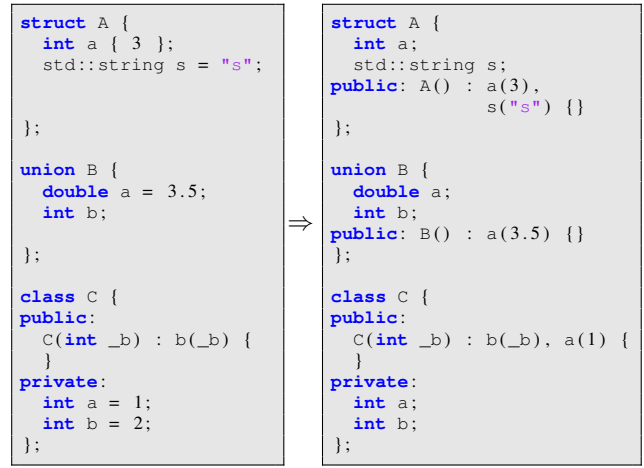

\centering
\begin{minipage}{0.41\linewidth}\centering
\begin{lstlisting}[frame=tlrb,breaklines=false]{Name}
struct A {
  int a { 3 };
  std::string s = "s";
  
  
};

union B {
  double a = 3.5;
  int b;
 
};

class C {
public:
  C(int _b) : b(_b) {
  }
private:
  int a = 1;
  int b = 2;
};
\end{lstlisting}%
\end{minipage}\codexformarrow\begin{minipage}{0.48\linewidth}\centering
\begin{lstlisting}[frame=tlrb,breaklines=false]{Name}
struct A {
  int a;
  std::string s;
public: A() : a(3),
              s("s") {}
};

union B {
  double a;
  int b;
public: B() : a(3.5) {}
};

class C {
public:
  C(int _b) : b(_b), a(1) {
  }
private:
  int a;
  int b;
};
\end{lstlisting}
\end{minipage}
\caption{In-class member initialization examples}
\label{lst:memberinit}
\end{figure}

The listing in Figure~\ref{lst:memberinit} shows examples for in-class member initialization.
The left-hand side of the figure lists the original C++11 code, and the other is the transformed version (C++03).
The mechanism used for the transformation is practically the one used by the compiler as well.
Namely, we move data initializers into the constructors provided they are not already present in the constructor initialization lists.
Automatically generated constructors need special consideration.
If they are not already generated by the front end, then our transformation framework will create them with public access specification (placed after the last existing member declaration in order not to accidentally modify visibility of other members).

Some member types are not handled by the framework because they cannot be transformed (or it is not practical) into its equivalent.
This includes C-style arrays, because their members cannot be directly initialized in the constructor initializer lists, only in the constructor bodies by individual value assignments.
Also, declarations in which multiple declarators are provided for the same type are not handled.
Finally, code is not transformed for template classes because in this case there might be constructors which are not instantiated by the front end, so consequently they could not be used to hold the generated code.

\begin{comment}
Ezen két esetre a példák \aref{lst:membunsup}. ábrán láthatóak.

\begin{figure}[htb]
\centering
\begin{minipage}{0.8\linewidth}
\begin{lstlisting}[breaklines=false]
class A {
  int a[3] = { 2, 3, 5};
  int b = 1, c = 2;
};
\end{lstlisting}
\end{minipage}
\caption{Not supported constructions}\label{lst:membunsup}
\end{figure}
\end{comment}

\subsection{Auto type deduction}

%\todoi{Arpi}

Prior to C++11, each variable (and other entity like a function return value) had to be explicitely declared for its static type.
In many cases, this led to overly complex and unreadable code.
The \texttt{auto} keyword used in place of a concrete type instructs the compiler to deduce the type of the entity automatically.
However, in this case, the variable needs to be intialized at the declaration in order the type be deducible.

Our transformation framework uses the same deduction rules as the compiler but in our case, the source code with the deduced types is generated as well.
In our implementation, various categories of auto types are distinguished, which is necessary because different treatments are required for the different cases:
\begin{itemize}[noitemsep]
\item simple declarations
\item multiple variables in one declaration
\item function pointers
\item template functions with such variables
\item functions with trailing return types
\end{itemize}

\begin{figure}[htb]
\centering
\begin{minipage}[htb]{0.44\linewidth}
\begin{lstlisting}[breaklines=false]
auto a = 32;
auto *b = new auto(&a);
auto xp = &a, yp = xp;
auto *y = &a, **z = &y;
auto foo(int a)
      -> decltype(a) {
  return a;
}
auto x = foo(0);
const auto & y = foo(1);
auto fp = foo;
\end{lstlisting}
\end{minipage}\codexformarrow\begin{minipage}[htb]{0.44\linewidth}
\begin{lstlisting}[breaklines=false]
int a = 32;
int **b = new int *(&a);
int * xp = &a, * yp = xp;
int * y = &a, ** z = &y;
int foo(int a) {

  return a;
}
int x = foo(0);
const int & y = foo(1);
int (*fp)(int) = foo;
\end{lstlisting}
\end{minipage}
\caption{Auto type deduction examples}
\label{lst:auto}
\end{figure}

The listing in Figure~\ref{lst:auto} shows examples for auto type deductions with original and transformed code versions.
This transformation has some limitations too.
Namely, multiple variables for a declaration in global scope, template functions, and certain variable declarations combined with preprocessor macros are not fully handled.

\begin{comment}
Bizonyos esetekben korlátozások bevezetésére volt szükség a keretrendszer helyes működésének megtartása érdekében.
Az egyik ilyen eset az, amikor globális láthatósági térben több változót deklarálunk egy kifejezésben.
Ezek az esetek nem kerülnek átalakításra.
A másik ilyen eset különböző preprocesszor direktívák használata esetén fordulhat elő, ahol az átalakítás után bizonyos kódrészek lefutása nem definiált.

\begin{figure}[ht]
\caption{Makró függő változó.}
\label{fig:preproc1}
\centering
\includegraphics[width=0.5\textwidth]{preproc.pdf}
\end{figure}

\Aref{fig:preproc1}.-es ábrán előforduló helyzetben a teljes transzformáció után, ha csak az \textbf{a.cpp}-t transzformáljuk újra nem lenne információnk arról, hogy más esetben a preprocesszor direktíva alkalmazása megtörtént-e, így az \texttt{\#ifdef} közé került kódot nem alakítanánk át.
A probléma megoldható, ha a \textbf{b.cpp}-t is újra transzformáljuk.
A jelenség azért fordulhat elő, mert keretrendszer csak a változott fájlokat alakítja át és ebben az esetben az eredeti fájlt és függőségeit átmásolja. 

\begin{figure}[ht]
\caption{Fájlokon átívelő függőség.}
\label{fig:preproc2}
\centering
\includegraphics[width=0.5\textwidth]{preproc2.pdf}
\end{figure}

\Aref{fig:preproc2}.-es ábrán látható esetben az \texttt{x} változó értéke függ attól, hogy előtte milyen fájl kerül \texttt{include}-olásra.
A transzformáció során az utolsó ilyen esetre lenne átalakítva a típusa. 
\end{comment}

\subsection{Lambda functions}
%\todoi{Arpi}

One of the most advanced new features in C++11 are lambda functions.
With them, special functionalities may be written inline in a very compact way, without actually creating new functions each time, and which was possible only using function pointers or function objects in previous editions of C++.
Our transformation engine translates lambda functions to function objects, as shown in the example in Figure~\ref{lst:lambda}.
%\todoi{valami reszlet azert kellene ide - Rudi: szerintem eleg ennyi -Arpi: hat, inkabb az attribute, final, override legyen rovidebb, pl pelda nelkul, hiszen az csak trivialis torlest jelent, emez meg szerintem az egyik legkomolyabb trafonk}

\begin{figure}[htb]
\centering
\vspace{-3mm}
\begin{minipage}[htb]{0.39\linewidth}
\begin{lstlisting}[frame=tlrb,breaklines=false]{Name}
std::vector<int> v(6);
int inc = 7;







std::for_each(
  v.begin(),
  v.end(),
  [&inc](int &n) {
    n += inc;
  }
);
\end{lstlisting}
\end{minipage}\codexformarrow\begin{minipage}[htb]{0.50\linewidth}
\begin{lstlisting}[frame=tlrb,breaklines=false]{Name}
std::vector<int> v(6);
int inc = 7;
class LambdaFunctor__12_1{
  int& inc;
public:
  LambdaFunctor__12_1(
    int& inc) : inc(inc) {}
  void operator()(int & n){
    n += inc;
  }
};
std::for_each()
  v.begin(),
  v.end(),
  (LambdaFunctor__12_1(inc))
);
\end{lstlisting}
\end{minipage}
\caption{Lambda function example}
\label{lst:lambda}
\vspace{-5mm}
\end{figure}

%A 2003-as szabványban egy olyan tiltás került bevezetésre, hogy a lokális osztályok nem lehetnek olyan függvények paraméterei, amelyek sablonparamétert várnak.
%Ez a megszorítás számos problémát vetett fel a fejlesztők körében és a legtöbb fordító nem is valósította meg ezt, mert túl nagy korlátozásnak tartották.
%Azonban a GNU GCC-ben beépítésre került ez a szigorítás, emiatt ezen fordító korábbi verzióinak támogatásához egy egyedi megvalósítást kellett létrehoznunk a függvényobjektumok használhatóságának  támogatására.

%A lambda kifejezések paramétereinek az átadása minden paraméter esetén külön szabályozható.
%A paraméterek lehetnek referenciaként vagy értékként átadva, továbbá módosító szavakat is megadhatunk.
%Keretrendszer támogatja ezeknek a paraméter típusoknak átalakítását is.

\subsection{Attributes}
%\todoi{Arpi}

%\todoi{ha helyszuke van, ez torolheto, hiszen az attributumok ugysincsennek tamogatva}

The reason of the introduction of attributes in C++11 was to unify the creation of various compiler directives.
Most compilers already implemented their dialect-specific ways for such directives, but this was not standard in any way (for example, construct like \texttt{\_\_attribute\_\_((...))} for GNU GCC and \texttt{\_\_declspec()} for the Microsoft compiler).
The use of attributes make this kind of extensions more portable, furthermore, they are very general and might be placed virtually at any syntactic position in the code, they might be placed in namespaces, can get parameters, etc.

\begin{figure}[h]
\vspace{-5mm}
\centering
\begin{minipage}{0.7\linewidth}
\begin{lstlisting}[breaklines=false]
[[attr1, attr2, attr3(args)]]
[[namespace::attr(args)]]
\end{lstlisting}
\end{minipage}
\caption{Attribute examples}
\label{lst:attrib}
\vspace{-2mm}
\end{figure}

Figure~\ref{lst:attrib} shows what kind of attributes are accepted by our transformation framework.
Since in the previous language versions there are no equivalent or similar code structures, we simply discard any occurrence of attributes from the code.

\subsection{Final and override modifiers}
%\todoi{Rudi}

%Az \texttt{final} és az \texttt{override} módosító szavak bevezetésének célja, hogy a fejlesztőknek támogatást adjon a fordító helyesség ellenőrzésnél.
The final and override modifiers were introduced to give developers compile-time control over class specialization and function overriding.
%Ezek a módosító szavak nem kulcsszóként jelentek meg a nyelvben, hanem környezettől függően lehetnek változó nevek is akár.
These modifiers are not keywords in the language, and depending on the environment they can appear also as e.g. variable names.
%Az \texttt{override} szót virtuális függvények esetén adhatjuk meg jelezve, hogy az ősosztály függvényét felüldefiniáltuk.
The \texttt{override} modifier indicates that the base class' virtual function is being overridden.
%A \texttt{final} szót használhatjuk virtuális függvények esetén vagy osztályok esetén is.
The \texttt{final} modifier can be used with both virtual functions and classes.
%Függvények esetében megtiltja a felüldefiniálást.
%Osztályok esetén pedig megtiltja, hogy az osztályt felhasználjuk öröklődés során ősosztályként.
In case of a function it prohibits its overriding, while in case of a class it disables subclassing.
%A keretrendszer ebben az esetben is csak törli a módosító szavakat, hasonlóan az attribútumokhoz. 
The framework simply deletes these modifiers, similarly as in case of attributes.
%\Aref{lst:finalover}. példában látható ezen módosító szavak alkalmazása osztályokon és tagfüggvényeken. A transzformáció utáni állapot pedig \aref{lst:finaloverafter}.-es ábrán látható.
The listing in Figure~\ref{lst:finalover} shows examples and their transformed versions.

\begin{figure}[htb]
\centering
\vspace{-3mm}
\begin{minipage}[htb]{0.47\linewidth}
\begin{lstlisting}[breaklines=false]
class A {
  virtual void b();
  virtual void c() final;
};
class B final : public A {
  void b() override final;
};
\end{lstlisting}
\end{minipage}\codexformarrow\begin{minipage}[htb]{0.36\linewidth}
\begin{lstlisting}[breaklines=false]
class A {
  virtual void b();
  virtual void c();
};
class B: public A {
  void b();
};
\end{lstlisting}
\end{minipage}
\caption{Final and override modifier examples}
\label{lst:finalover}
\vspace{-5mm}
\end{figure}

\subsection{Range-based for loop}
%\todoi{Rudi}

%Annak érdekében, hogy a \texttt{for} ciklus alkalmazását megkönnyítsék olyan esetekben, amikor egy teljes tartományon szeretnénk műveleteket végrehajtani, bevezettek egy kompaktabb írási módot.
In order to use the \texttt{for} loop easier in cases where an operation has to be performed on a whole range of elements, a more compact way of writing code was introduced.
%Amennyiben az adott objektum rendelkezik a szükséges kitüntetett függvényekkel, akkor használható ez az egyszerűsített forma.
If the given container object has all the required special functions, it can be used in this simplified form.
%Ezek a kitüntetett függvények a \texttt{begin} és az \texttt{end}.
These special functions are called \texttt{begin} and \texttt{end}.
%Kivételt képeznek az egyszerű tömbök ezen feltétel alól, azonban ott memória cím eltolással meghatározható ezen tartomány.
An exception from this requirement are simple arrays, because in this case the range can be determined by calculating memory address offset.
%A meghatározott függvények lehetnek globálisak vagy lokálisak.
The special functions can be global or local.
%Ha a kettő kitüntetett metódus a típus deklarációban van definiálva paraméter nélkül, akkor lokálisnak nevezzük, ha a globális névtérben található és az objektumot várja paraméterként, akkor globálisnak nevezzük.
They are local if the two methods are defined in the class declaration and have no parameters, and global if they are defined outside the class in its enclosing namespace and have a parameter of the required class type.

\begin{figure}[htb]
\centering
\vspace{-5mm}
\begin{minipage}[htb]{0.44\linewidth}
\begin{lstlisting}[breaklines=false]
int array[4]={1,2,3,0}; 


for (auto &k : array) {


  k = 1;
}
\end{lstlisting}
\end{minipage}\codexformarrow\begin{minipage}[htb]{0.44\linewidth}
\begin{lstlisting}[breaklines=false]
int array[4]={1,2,3,0}; 
int * __begin1 = (array);
int * __end1 = (array)+4;
for(;__begin1 != __end1;
      ++__begin1) {
  int &k = *__begin1;
  k = 1;
}
\end{lstlisting}
\end{minipage}
\caption{Range-based for loop example}
\label{lst:forafter}
\end{figure}

%Az átalakítás során az új kompaktabb ciklus szintaxis visszaalakításra kerül a több paraméterrel meghatározható számláló ciklus szintaxisára, ahogyan \aref{lst:forafter}. példán is látható.
During the transformation the new compact syntax is converted to the old form with three arguments as shown in the example code in Figure~\ref{lst:forafter}. Note that the introduced local variables are suffixed with a number to avoid name clash with further
transformations in the same scope.

\subsection{Constructor delegation}
%\todoi{Rudi}

%Az úgynevezett konstruktor delegálásra is lehetőséget biztosít a C++11.
C++11 allows the delegation of constructors.
%Ezáltal a konstruktorok inicializációs listájában már megadható egy másik konstruktor is.
This means that in the constructor initialization list another constructor can be called.
%Ekkor csak ez az egyetlen eleme lehet a listának adott konstruktornál, ahogyan \aref{lst:delegbefore}. példában is látható.
In this case the constructor initialization list can contain only this single element.
%A konstruktor delegálás segítségével egyszerűbbé válik az objektumok inicializálása.
By using constructor delegation lots of copied code can be avoided when several constructors would perform similar initializations.

\begin{figure}[htb]
\centering
\vspace{-5mm}
\begin{minipage}[htb]{0.44\linewidth}
\begin{lstlisting}[breaklines=false]
class A {
  A() {}
  A(string str) : s(str)
  {
    t = "hello";
  }
  A(string str, int dbl)
    : A(str) {

    a = dbl;
  }
  int a = 1;
  string s;
  string t;
};
\end{lstlisting}
\end{minipage}\codexformarrow\begin{minipage}[htb]{0.44\linewidth}
\begin{lstlisting}[breaklines=false]
class A {
  A() : a(1) {}
  A(string str) : s(str),
                  a(1) {
    t = "hello";
  }
  A(string str, int dbl)
      : a(1), s(str) {
    { t = "hello"; }
    a = dbl;
  }
  int a;
  string s;
  string t;
};
\end{lstlisting}
\end{minipage}
\caption{Constructor delegation example}
\label{lst:delegafter}
\end{figure}

%Keretrendszer a delegáló konstruktorokat úgy alakítja vissza, hogy a delegált konstruktorok inicializációs listáját és függvénytörzseit a jelenlegi konstruktor listájába illetve függvénytörzsébe másolja.
%\Aref{lst:delegafter}. példa jól szemlélteti az átalakítást.
The framework transforms the code in such a way that it copies the initialization list of the target constructor into the initialization list or body of the caller constructors, as can be seen in Figure~\ref{lst:delegafter}.
%Amennyiben a delegáló konstruktorok sablonosztályban találhatók, akkor a keretrendszer csak a használt függvényeket tudja átalakítani.
If the constructor delegation is used in template classes then the framework can transform only the instantiated constructors.

\subsection{Type aliases}
%\todoi{Rudi}

%A C++ nyelvben már régóta támogatott típusok újabb névvel történő ellátása a \texttt{typedef} kulcsszóval.
%Régi megoldás nem támogatta azt, hogy olyan elnevezéseket készítsünk, amelyek fogadhatnak sablon paramétereket.
Supporting typedef-names is a long-standing feature of C and C++ to create aliases for existing types, but it does not support aliases which can receive template parameters.
%A C++11 ezen hiányosságra nyújt megoldást új nevezési szintaxis bevezetésével.
C++11 introduced a new syntax to support this feature with the \texttt{using} keyword.
%Az új szintaxis egyértelműbb formátumot nyújt az elnevezések létrehozásánál, valamint fogadhat sablon paramétereket is.
%Sablon paraméterek használata jól jöhet sablonosztályok átnevezése esetén.
Using template parameters can come in handy in case of creating aliases for template classes.
%Az új formátum alkalmazása \ref{lst:ta}. példán látható.
The listing in Figure~\ref{lst:ta} shows an example.

%A keretrendszer az új szintaxist a régi formátumra konvertálja az egyszerű esetben.
The framework converts the new syntax into the old format in simple non-template cases in a straightforward way.
%A sablon paraméteres verzió előfordulása esetén egy struktúrává alakítja az elnevezést, aminek egy típuselnevezési adattagja érhető el, ezáltal biztosítva a kompatibilitást a 2003-as szabvánnyal.
When there are template parameters, it creates a struct carrying the alias name and it inserts a typedef with the name `type' into it. Also, all references to the alias are replaced by this construct.
The listing in Figure~\ref{lst:ta} shows the transformed example code.
Occurrences of the alias name in symbol import statements (\texttt{using} from base class, for example), and dependent names
as alias parameters (requiring \texttt{typename} prefix for the nested type) are currently not supported.

\begin{figure}[htb]
\centering
\vspace{-7mm}
\begin{minipage}[htb]{0.44\linewidth}
\begin{lstlisting}[breaklines=false]
using ul = unsigned long;
ul foo(ul p) {return p;}

template<class T> 
using mapVec=std::map
  <T, Vec<T> >;



mapVec<int>
bar(mapVec<int> p) {
  return p;
}
\end{lstlisting}
\end{minipage}\codexformarrow\begin{minipage}[htb]{0.46\linewidth}
\begin{lstlisting}[breaklines=false]
typedef unsigned long ul;
ul foo(ul p) {return p;}

template<class T> 
struct mapVec {
  typedef std::map
    <T, Vec<T> > type;
};

mapVec<int>::type
bar(mapVec<int>::type p) {
  return p;
}
\end{lstlisting}
\end{minipage}
\caption{Type alias examples}
\label{lst:ta}
\vspace{-3mm}
\end{figure}

\section{Evaluation}
\label{chap:eval}

We evaluated our transformation framework from two aspects: correctness of the transformed code and performance (runtime).
The first aspect is clearly important since we want the transformed code be functionally equivalent to the original one.
However, note that there are language constructs that are not handled by the framework, so these were excluded from our measurements (and were, of course, communicated to the users).
We discuss functional testing in Section~\ref{chap:eval:test}.

The second aspect of the evaluation, performance testing, is important since the framework is planned to be used in production by our industrial partner, integrated into the build process.
Since the company employs frequent builds, which is resource intensive due to the large and complex code base, time to perform the transformation is also critical.
Associated measurements are provided in Section~\ref{chap:eval:perf}.

%\subsection{Subject systems}
%\todoi{Arpi}

During development and early stages of the evaluation, we used a set of code snippets with the language features of interest.
Later we relied on a benchmark of systems, which use some of the C++11 features, and are non-trivial in size.
We included two kinds of systems: four open-source systems and two proprietary ones.
Some basic properties of the subject systems are provided in Table~\ref{eval:apps}.
All subjects belong to different domains, and the sizes of the open-source systems range from small to medium, while the industrial ones can be treated as large systems.
The first industrial system is Columbus, our own source code analysis framework~\cite{FBT02}.
The other system is iGO, the product of our industrial partner NNG, which was the initial motivation for this work.

\vspace{-3mm}

\begin{savenotes}
\begin{table}[htb]
\centering
%\small
\caption{Properties of the subject systems}
\label{eval:apps}
\begin{tabular}{lrrr}
\toprule
            & LOC & Transl. units & Transformations \\
\hline
SoDA\footnote{https://github.com/sed-szeged/soda}
        &    18,849    &       126   &       193 \\
log4cplus\footnote{https://github.com/log4cplus/log4cplus}
        &    37,543    &        67   &       172 \\
GridDB\footnote{https://github.com/griddb/griddb\_nosql}
        &   113,270    &        68   &        13 \\
aria2\footnote{https://github.com/aria2/aria2}
        &   118,063    &       385   &     3,388 \\
\hline
Columbus
        &   889,725    &      1,462   &       343 \\
iGO
        &  millions\footnote{the exact figure is confidential} &       121   &       0 \\
\bottomrule
\end{tabular}
%\todoi{footnote és table nem barátok}
\end{table}
\end{savenotes}

Lines of Code (LOC), given in the second column is counted as logical lines (not including empty and comment lines), while the number of translation units is essentially the number of source files with extension \texttt{.cpp}, that are compiled by the compiler during build.
The last column of the table shows the number of transformations performed by the system during the whole process.
It can be observed from the statistics that the actual number of transformed language elements varied from program to program and it did neither really correlate with program size nor with the number of translation units.

The reason behind the surprisingly low number of compilation units in the iGO system is a build time optimization technique called \emph{unity build}.
It works by processing a set of compilation units together so that multiple redundant processing of header files is radically reduced~\cite{jocophd}.
For iGO, there were no actual transformations performed, which is discussed in the following.

%A táblázatban található kettő IPARI jelölésű projekten is sikeresen lefutott az alkalmazásunk, azonban itt nem áll rendelkezésünkre részletes mérési eredmény és az egyik esetben C++11-es nyelvi elemek nem találhatóak, mert ezen rendszer volt a motivációja az egész keretrendszernek.

\subsection{Functional testing}
\label{chap:eval:test}

%\todoi{Arpi}
The correctness of the transformed code was checked in two steps.
First, the transformation framework is capable of checking if the code is syntactically correct, so after each successful transformation this check was also performed.
Second, the code has to produce the same behavior as the original one, and this property was verified at multiple levels:
\begin{enumerate}
\item We wrote a set of code snippets containing examples of the implemented transformations (see Table~\ref{tab:tests} for their amount).
These pieces of test code have been transformed, syntactically checked, and compiled in the legacy environment.
Then, each example was manually verified, and finally executed on one or two test cases for functional equivalence.
These tests are part of the transformation framework available open-source.
\item On the four open-source systems and Columbus we also performed the transformation, syntax check, and legacy compilation.
Finally, we manually verified a limited number of transformations performed in these systems (due to their large number, we could not check all).
\item In the case of iGO, there were no actual transformations performed, as can be observed from Table~\ref{eval:apps} as well.
This is because at the time of the experiments the code base did not include any C++11 features.
However, the other parts of the process -- analysis, compilation, integration into the build process, incrementality, etc. --  were verified.
To check the actual working of the transformation engine, the code was temporarily modified at a few places to include C++11 code.
\end{enumerate}

Despite the fact that no actual transformation has been done on iGO yet, the above functional testing process ensures future usability of the framework on this system as well.
The transformed code needed to be platform independent, so we performed the tests on Windows and Linux environments with different compiler versions as well.

\vspace{-3mm}

%\todoi{igo-n megsem 0 trafo volt?!}

\begin{table}[ht!]
\centering
\caption{Code snippets for functional testing}
\label{tab:tests}
\begin{tabular}{lr}
\toprule
Transformation            & Code snippets \\
\hline
In-class data member initialization &  3 \\
Auto type deduction                 & 37 \\
Lambda functions                    & 31 \\
Attributes                          &  3 \\
Final and override modifiers        &  3 \\
Range-based for loop                &  9 \\
Constructor delegation              &  2 \\
Type aliases                        &  3 \\
\hline
All                                 & 91   \\
%attribute                & 2             \\
%decltype                 & 8             \\
%deduce-auto              & 20            \\
%delegating-constructor   & 2             \\
%final-override           & 3             \\
%instantiate-template     & 5             \\
%macro                    & 4             \\
%%move                     & 5             \\
%range-based-for          & 9             \\
%replace-lambda           & 30            \\
%replace-member-init      & 3             \\
%template-metaprogramming & 2             \\
%type-alias               & 3             \\
%%variadic                 & 5             \\
\bottomrule
\end{tabular}
\end{table}

\begin{comment}
Egy tesztesetet akkor tekintünk sikeresnek, ha az alábbi feltételek mindegyike teljesül:
\begin{itemize}[noitemsep]
\item hiba nélküli transzformáció és szintaktikai ellenőrzés a keretrendszerrel,
\item alkalmazás fordítása egy választott fordítóval,
\item alkalmazás hiba nélküli sikeres futtatása az adott platformon.
\end{itemize}
\end{comment}

\vspace{-3mm}

\subsection{Performance testing}
\label{chap:eval:perf}

In order to improve the applicability of our framework on big systems we implemented different speedup techniques to reduce the overall processing time:
\begin{itemize}[noitemsep]
\item Transformation is running in multiple threads in \emph{parallel}.
\item \emph{Incremental} transformation performs only the necessary steps based on what has changed since the last transformation.
\item \emph{Feature finder} identifies what language features are used in the different compilation units to eliminate their superfluous processing in the unrelated transformation rounds.
\item The \emph{MultipleTransforms} phase performs transformations of certain independent language features in a single round.
\end{itemize}
The following discussion presents measured processing times and other empirical results on our reference code bases.
%The numbers for the iGO system are confidential, so we had to exclude it from this set of results.

\begin{savenotes}
\begin{figure}[htb]
\centering
\includegraphics[width=\linewidth]{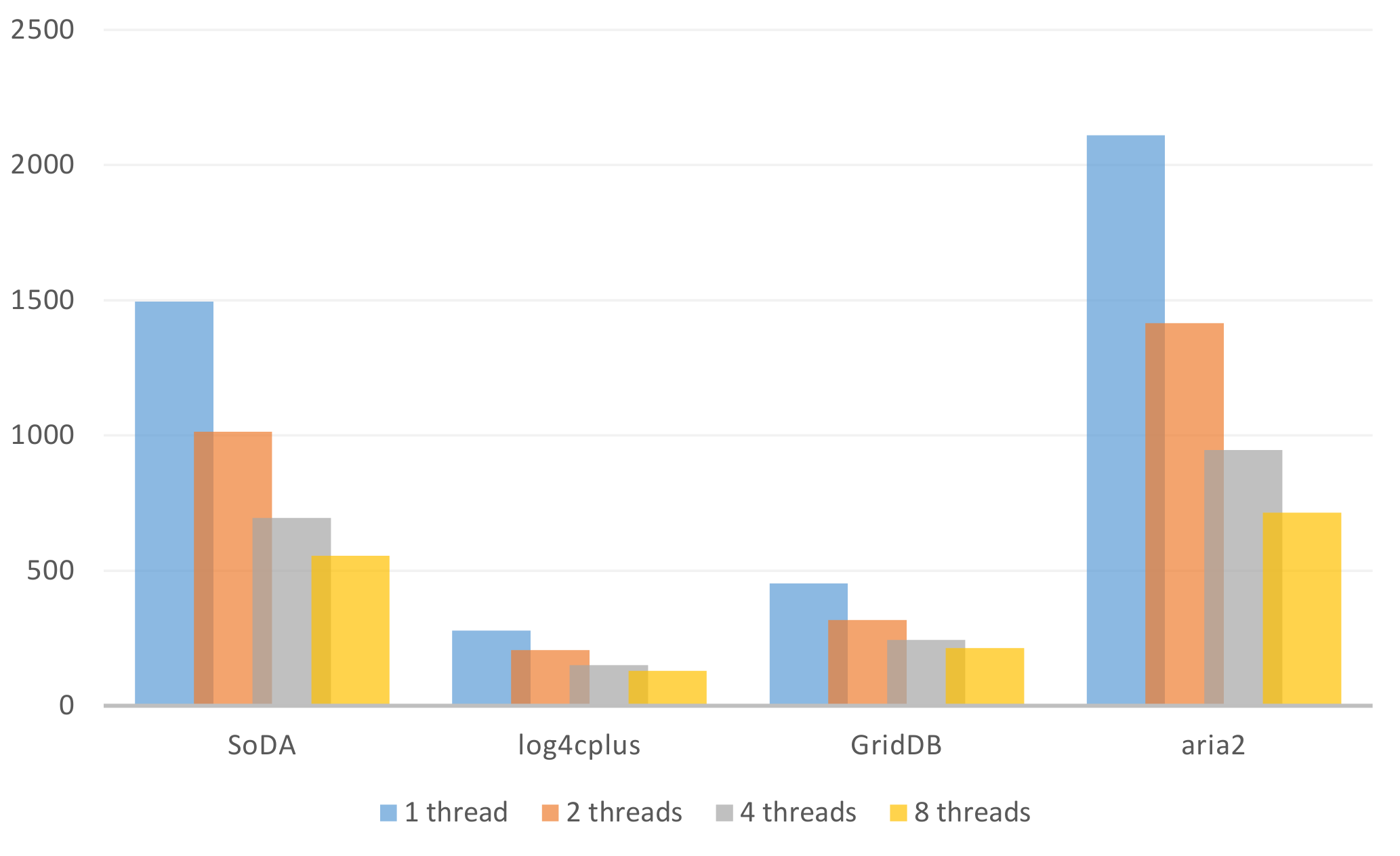}
\vspace{-6mm}
\caption{Runtime in seconds.}
\label{fig:parallelism}
\vspace{-6mm}
\end{figure}
\end{savenotes}

Figure~\ref{fig:parallelism} shows total processing times on the code bases of the four open-source systems in seconds.\footnote{Source code was accessed via a mapped network drive, presumably resulting in slower than usual file access times, somewhat distorting the measurements.}
The same performance test was performed with different parallelization settings (how many threads to use) to determine the scalability of the framework.
Note that \textit{GridDB} has a big advantage in terms of translation time compared to \textit{SoDA}, even though the LOC measure of the former is 6 times of that of the latter.
The big difference is caused by not including 3rd party code when counting LOC, while the transformation framework has to analyze 3rd party code as well.
Systems may have certain large 3rd party codes embedded into their own code base, resulting in lots of extra instructions to process by the transformation framework.
Currently the last phase, when syntax check is performed, does not support parallel execution, which reduces scalability to multiple cores.

\begin{table*}[htb]
\centering
\caption{Detailed runtime data without parallelization (in seconds)}
\vspace{-3mm}
\label{eval:details}
\resizebox{\textwidth}{!}{
\begin{tabular}{l|rrrrrrr}
\toprule
  & Dep. analysis & FeatureFinder & ReplaceLambda & MultipleTransforms & RemoveAutoDelegation & Syntax check & Total \\
\hline
SoDA        &  47 &   853 &  3 &   281 &  35 &   239 & 1,458 \\
log4cplus   &   3 &   136 &  3 &    69 &  12 &    48 &   271 \\
GridDB      &   4 &   285 &  0 &    57 &   0 &   103 &   449 \\
aria2       &  20 &   860 & 16 &   508 & 222 &   333 & 1,959 \\
\hline
Columbus    & 142 & 4,631 & 73 & 2,672 & 617 & 1,345 & 9,480 \\
% iGO         & 201  & 2,527 & N/A & N/A & N/A & N/A \\
iGO         & 202 & 2,319 & N/A & N/A  & N/A &   905 & 3,426 \\
\bottomrule
\end{tabular}
}
\vspace{-3mm}
\end{table*}

%\todoi{az egyes fazisok itt derult egbol vannak felsorolva}

Table~\ref{eval:details} presents processing times by phases without parallelization.\footnote{Although iGO did not contain any C++11 features to transform, the other phases of the process were executed.}
%\todoi{A fazisokat jo lenne roviden ismertetni.}
%\todoi{A tablazat vegen legyen egy Total oszlop is, ami egyebkent meg kellene hogy egyezzen a korabbi adatokkal.}
The transformation starts with \textit{Dependency analysis} which checks each compilation unit and its dependencies and decides whether the compilation unit has to be transformed or not.
The most time consuming phase is clearly shown to be \textit{FeatureFinder}, being responsible for identifying language feature usages, because it has to examine all compilation units.
The transformation phases (\textit{ReplaceLambda}, \textit{MultipleTransform} and \textit{RemoveAutoDelegation}) and Syntax check phase (which verifies the transformed code) only deal with units containing code fragments relevant to the actual transformation phase.
The big differences between times of \textit{FeatureFinder} and the certain transformation phases reveal how much time is saved by the feature finder optimisation.
Though transformation phases do not only parse source code, but also transform it, time spent in actual code transformations was measured to be negligible compared to parsing time.

\begin{figure}[htb]
\centering
\includegraphics[width=\linewidth]{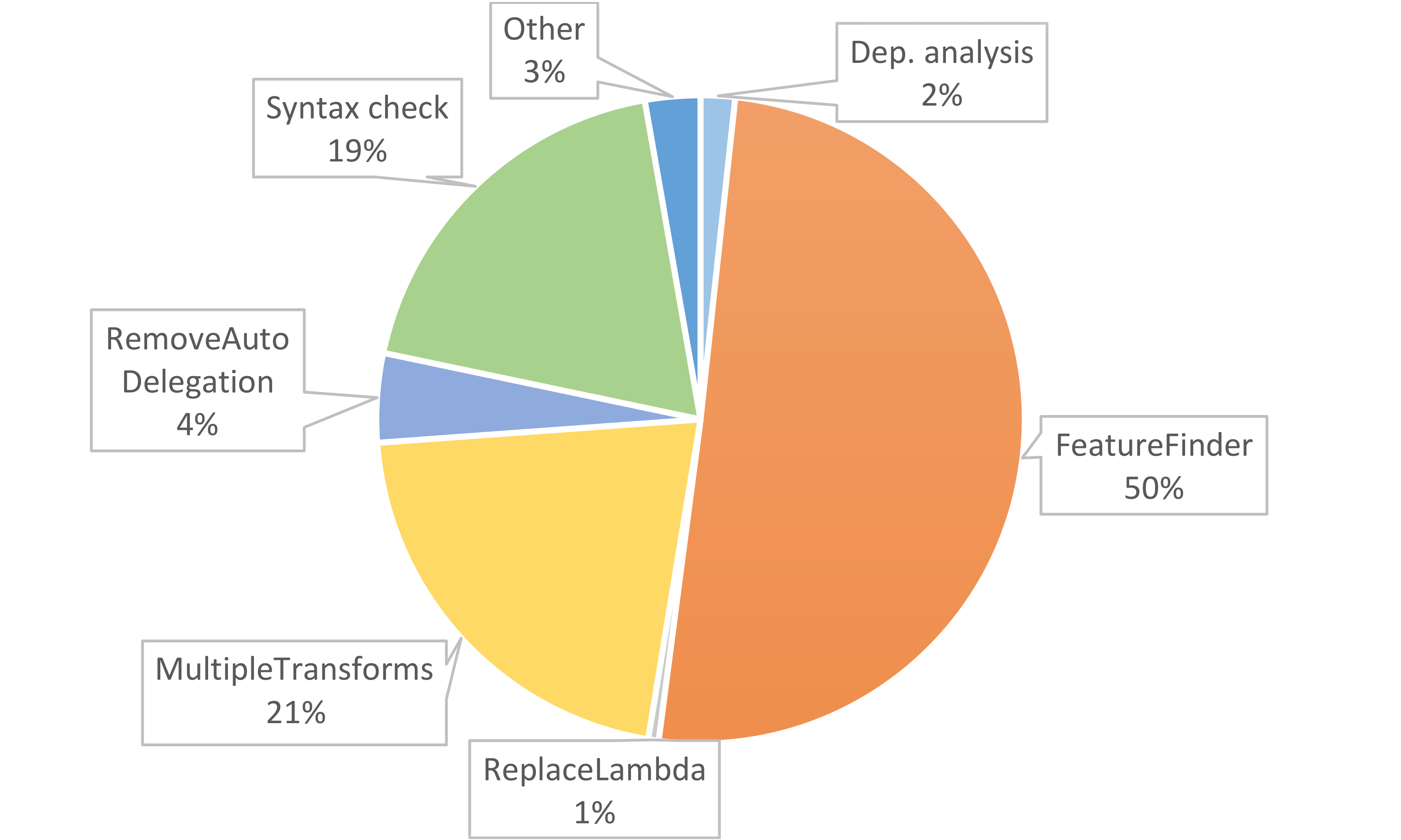}
\caption{Average distribution of time spent in transformation phases}
\label{fig:evaldist}
\end{figure}

The distribution of the four open-source systems' processing time among the different transformation phases is shown in Figure~\ref{fig:evaldist}.
The numbers were determined by averaging values of Table \ref{eval:details}.
71\% of the time is spent in the \textit{FeatureFinder} and \textit{MultipleTransforms} phases.
Without \textit{FeatureFinder} the distribution would probably be more equalized, since each phase would contain very similar parsing and the negligible code transformation steps for the same complete set of compilation units.
\textit{MultipleTransforms} eliminates entire transformation phases by uniting the processing of independent language features.
%\todoi{miert is lenne azonos?}

\section{Limitations}
\label{chap:limit}

%\todoi{0.5 oldal Arpi, inkabb legyen uj section szerintem}

%\subsection{További transzformációk}

%\todoi{az 5-os lista ezzel kiegeszulve vajon kiadja-e a Table III eseteit?}

Apart from the ones listed in Section~\ref{chap:trafo}, our framework implements several other transformations, though with limited functionality.
This includes the following language features: variadic templates, rvalue references, move constructors and \texttt{decltype} specifiers used for type deduction.
These features can be used provided some constraints are met by the developers, but since the most typical usage scenarios are handled, this does not mean serious limitation in practice.

In Section~\ref{chap:trafo}, we already listed some concrete limitations for the transformations (e.g., 
%type deduction through preprocessor macros, 
unused template methods, deletion of attributes).
Apart from these, if the framework encounters some specific variants of language features that are not fully handled, it tries to skip those parts and continue the analysis, before eventually terminating with an error.
If the system contains code that is generated during compilation, the framework will not consider these files.

Fully automatic generation of the \texttt{compile\allowbreak\_\allowbreak commands\allowbreak.json} file required for building with the clang infrastructure is not supported.
In Linux, the CMake\footnote{https://cmake.org} system provides functionality for generating this file, while on other systems Bear\footnote{https://github.com/rizsotto/Bear} might be used.
However, some additional modifications are needed to be made on the generated file in order to be compatible with the transformation framework.
As far as we know, for Windows systems there is no universal solution for producing the build file, so in this case the user has to provide it.
A particular issue on Windows is related to older Visual Studio versions,\footnote{https://msdn.microsoft.com/en-us/library/ms950416.aspx} in which case the project file has to be prepared (or updated) in multiple versions, one for each Visual Studio edition.

Finally, each subject system to be transformed needs to be compilable by the clang compiler, because this is what our framework is built on.
Systems not satisfying this property might require significant porting effort before being capable of transformation.

\section{Conclusion}
\label{chap:concl}

%\todoi{1 oldal referenciakkal, Rudi}

%Először egy általános körképet mutattunk a vállalati partnerek esetén felmerülő problémákról.
%Majd részletes indoklást adtunk arról, hogy miért is volt szükség az általunk fejlesztett rendszernek a kivitelezésére. 
%We started this paper by showing a general overview of the problems companies are facing when they need to produce C++03 code but their developers are eager to use the new features of C++11.
There are many reasons why companies are facing problems when they need to produce C++03 code but their developers are eager to use the new features of C++11.
%Részletesen ismertettünk különböző transzformációs rendszereket és támogató könyvtárak, amelyeket forráskód transzformációs műveletekre használhatunk.
This motivated our work to construct a system for automatically transforming C++11 code to C++03.
%A rendszer megadja azt a lehetőséget, hogy bizonyos megszorítások mellett a fejlesztők használni tudjanak különböző C++11-es nyelvi elemeket úgy, hogy a rendszerük az átalakítás után továbbra is kompatibilis lesz a régebbi C++03-as szabvánnyal.
The system allows, under certain restrictions, for developers to use various C++11 language elements so that after conversion, software will continue to be compatible with the older C++03 standard.
%Párhuzamot vontunk a mi keretrendszerünk és más hasonló rendszereket között.
%We drew a parallel between our framework and other similar systems.
%Összegzésképpen, a korábbi fejezetekben bemutattunk egy olyan új transzformációs keretrendszert, amely C++11-es nyelvi elemeket képes C++03-as szabvánnyal kompatibilis alakra transzformálni.
%In summary, we introduced a new transformation framework that is able to transform C++11 language elements to C++03 compliant source code.
%A rendszer úgy lett megtervezve, hogy jól integrálható legyen a legkülönfélébb fejlesztési folyamatokba, ezért számos egyéb szolgáltatást is biztosít, mint például az inkrementális átalakítás, forráskód struktúra klónozás és a forráskód visszakövethetőség.
We designed the system in a way that it can be easily integrated into a wide range of development processes.
In addition, it provides several other services, such as incremental transformation, cloning source code structure, and source traceability.

%Ezek után az általunk fejlesztett rendszer tulajdonságait és képességeit részleteztük.
We detailed the features and capabilities of our source to source transformation system,
%Először a megvalósított keretrendszer alapvető képességeit és a felépítését mutattuk be.
which includes the basic structure and operation of the framework, 
%Majd a következőkben részletesen, példákon keresztül a megvalósított transzformációkat mutattuk be, valamint néhány kísérleti fázisban levő átalakítást is ismertettünk.
the implemented transformations with examples, and information on how we tested them.
%Végezetül a rendszer teljesítményét értékeltük ki különböző nyílt forráskódú alkalmazásokon futtatva, kiemelve a skálázhatóságát és néhány a kísérletek során tapasztalt limitációkat adtunk közre.
We evaluated system performance on different open-source applications and on two large industrial systems, highlighting the scalability and some limitations we encountered.
We know that the testing methodology we used for validation could be enhanced further, but current experience shows that the method is already usable in practice.

%Jövőbeli terveink között szerepel a rendszer nyílt forráskódúvá tétele.
The developed framework is open-source and it can be freely used.
%Ezen felül számos további fejlesztési lehetőség kínálkozik a rendszer továbbfejlesztésére.
%Új nyelvi elemek átalakítását lehet kivitelezni, a jelenlegi transzformációk hibáját korrigálni.
%A szintaxis ellenőrző párhuzamos futásának a megvalósítása.
%Valamint a hibaelkerülő mechanizmuson is lehet fejlesztéseket végezni.
There are many opportunities for further development, however.
For instance, handling new language elements, correction of current transformation errors, and the improvement of error recovery mechanisms.

\vspace{-1mm}

\section*{Acknowledgment}

We are grateful to our industrial partner for the interesting project.
The authors would like to thank \'Ad\'am Mar\'oti, \'Ad\'am Rudas and K\'aroly Szab\'o for their work in the implementation.
This work was partially supported by the European Union FP7 project REPARA (no: 609666). 

\vspace{-1mm}

\bibliographystyle{plain}
\bibliography{references}

\end{document}